\documentclass[aps,prd,epsf]{revtex4}

\begin{document} 

\title{
Boundary terms for eleven-dimensional supergravity and $M$-theory} 
\author{Ian G. Moss}
\email{ian.moss@ncl.ac.uk}
\affiliation{School of Mathematics, University of  
Newcastle Upon Tyne, NE1 7RU, UK}

\date{\today}


\begin{abstract}
A new action for eleven dimensional supergravity on a
manifold with boundary is presented. The action is a possible low energy limit
of $M$-theory. Previous problems with infinite constants in the action are
overcome and a new set of boundary conditions relating the behaviour of the
supergravity fields to matter fields are obtained. One effect of these boundary
conditions is that matter fields generate gravitational torsion.  
\end{abstract}

\maketitle

One of the phenomenologically interesting low energy limits of $M$-theory
corresponds to eleven-dimensional supergravity on a manifold with a
ten-dimensional boundary. Horava and Witten have argued that this theory gives
the strongly coupled limit of the $E_8\times E_8$ heterotic string
\cite{horava96,horava96-2}. The gauge and matter fields propagate on the
boundary and couple to the the bulk supergravity. Previous attempts to
construct these couplings have been hampered by the appearance of the square
of the Dirac delta function. This paper presents an improved construction
which results in a consistent set of interaction terms.

The interaction terms in the model are constructed as an expansion in
$\kappa^{2/3}$, where $\kappa$ is the eleven dimensional gravitational coupling
constant. At leading order, the theory is eleven-dimensional supergravity on
the background $R^{10}\times S^1/Z_2$. Since the orbifold $S^1/Z_2$ is
identical to an interval $I$, with the covering space $S^1$, the background
has a boundary consisting of two timelike ten-dimensional surfaces or branes.
One of these contains the visible matter with which we are familiar and the
the other contains hidden matter. Gravitational forces are transmitted through
the interior which is often called the bulk.

The modified theory described here is presented from the point of view of the
manifold with a boundary, rather than the covering space. The action allows
for the possibility that the boundary may curve, as we would require for
descriptions of cosmology. The main inovations appear in the boundary
conditions. In the first place there are corrections to the chirality condition
on the gravitino, which are to a large extent forced by the requirements of
supersymmetry. There are also boundary conditions for the supergravity
three-form which play an important role. The resulting theory is
supersymmetric at orders $\kappa^0$, $\kappa^{2/3}$ and $\kappa^{4/3}$.

The notation used here is based on \cite{berg82}. The metric signature is
$-+\dots+$. Vector indices are $I,J,\dots$ in the bulk, $A,B,\dots$ on the
boundary and $N$ in the (unit) normal direction. The outward going normal has
extrinsic curvature $K_{AB}$. The exterior derivative of an $n$-form $\alpha$
is $({\rm d}\alpha)_{I_1\dots I_
{n+1}}=(n+1)^{-1}\partial_{[I_1}\alpha_{I_2\dots
I_{n+1}]}$ and the wedge product $(\alpha\wedge \beta)_{I_1\dots
I_m}={}_mC_n\,\alpha_{[I_1\dots I_n}\beta_{I_{n+1}\dots I_m]}$ where ${}_mC_n$
is a binomial coefficient. 
The gamma matrices satisfy $\{\Gamma_I,\Gamma_J\}=2
g_{IJ}$ and $\Gamma^{I\dots
K}=\Gamma^{[I}\dots\Gamma^{K]}$. The spinors are Majorana, and
$\bar\psi=\psi^T\Gamma^0$.

The action will be presented first, followed by an account of the boundary
conditions. The supersymmetry of the action will be discussed at the end. The
supergravity part contains the metric $g$, gravitino $\psi_I$ and three-form
$C$ (with field strength $G=6\,{\rm d}C$) \cite{cremmer78}. The usual
supergravity action is
\begin{eqnarray}
S_{SG}=&&{2\over \kappa^2}\int_{\cal M}d\mu\left(-\frac12R
-\frac12\bar\psi_I\Gamma^{IJK}D(\Omega^*)_J\psi_K-\frac1{48}G_{IJKL}G^{IJKL}\right.
\nonumber\\
&&\left.-\frac{\sqrt{2}}{192}\left(\bar\psi_I\Gamma^{IJKLMP}\psi_P
+12\bar\psi^J\Gamma^{KL}\psi^M\right)G^*_{JKLM}
-\frac{\sqrt{2}}{10!}
\epsilon^{I_1\dots I_{11}}(C\wedge G\wedge G)_{I_1\dots I_{11}}\right)
\end{eqnarray}
In the 1.5 order formalism used here, the spin connection $\Omega$ is varied as
an independent field. The combination
$\Omega^*=(\Omega+\hat\Omega)/2$, where hats denote a standardised subtraction
of gravitino terms to make a supercovariant expression.

The background for the low energy limit of $M$ theory of interest here contains
a manifold with timelike boundaries. The boundary conditions are fixed by
symmetries on the covering space. On a manifold with boundary, the Einstein
action is supplimented by an extrinsic curvature term
\cite{york72,gibbons77,chamblin99}.
The supersymmetric version has been found previously \cite{luckock89}. The full
boundary term with the extrinsic curvature and supersymmetry corrections is
\begin{equation}
S_0={2\over \kappa^2}\int_{\cal\partial M}d\mu\left(
K-\frac14\bar\psi_A\Gamma^{AB}\psi_B+\frac12\bar\psi_A\Gamma^A\psi_N\right)
\end{equation}
Variation of the action $S_{SG}+S_0$ gives a boundary condition on the
extrinsic curvature and a chirality condition on the gravitino $P_+\psi_A=0$,
where $P_+=(1+\Gamma_N)/2$. These boundary
conditions are are consistent with the symmetries of the orbifold $S^1/Z_2$.
The Bianchi identity implies an additional boundary condition $G_{ABCD}=0$
which cannot be derived from the action.

The gauge multiplet contains an $E_8$ gauge field $A^a_A$ and chiral fermions
$\chi^a$ in the adjoint representation. 
The action is based on the Yang-Mills action with an interaction term involving
the supercurrent,
\begin{equation}
S_1=-{2\epsilon\over\kappa^2}\int_{\cal \partial M}d\mu
\left(\frac14{F^a}_{AB}{F^a}^{AB}+\frac12\bar\chi^a\Gamma^AD_A(\hat\Omega)\chi^a
+\frac14\bar\psi_A\Gamma^{BC}\Gamma^A{F^a}^*_{BC}\chi^a
+\frac1{192}\bar\chi\Gamma_{ABC}\chi\bar\psi_D\Gamma^{ABCDE}\psi_E\right)
\end{equation}
The constant $\epsilon$ sets the relative scale of the matter coupling. The
total action $S=S_{SG}+S_0+S_1$. Note that $F^*=(F+\hat F)/2$, where
\begin{equation}
\hat F^a{}_{AB}=F^a{}_{AB}-\bar\psi_{[A}\Gamma_{B]}\chi^a.
\end{equation}
The complicated four fermi interaction in the action is required for obtaining
supercovariant boundary conditions.

One boundary condition is imposed as a constraint. The tangential components of
the three form will be required to satisfy
$c_{ABC}=0$, where
\begin{equation}
c_{ABC}=C_{ABC}+\frac{\sqrt{2}}{12}\epsilon\,\omega_{ABC}+
\frac{\sqrt{2}}{48}\epsilon\,\bar\chi^a\Gamma_{ABC}\chi^a\label{cbc}
\end{equation}
This boundary condition contains the Chern-Simons form,
\begin{equation}
\omega={\rm tr}\left(A\wedge dA+\frac23A\wedge A\wedge A\right).
\end{equation}
We shall see shortly how this boundary condition is fixed by supersymmetry, but
first it is interesting to check consistency with the gauge symmetry. Under a
non-abelian gauge transformation with $\delta A^a_A=-D_A\varepsilon^a$, the
variation in the Chern-Simons form becomes $\delta\omega=d(\varepsilon^aF^a)$.
The non-abelian gauge transformation of the Chern-Simons form can therefore be
absorbed by an abelian gauge transformation of the three-form (in analogy with
Yang-Mills Supergravity \cite{chapline83}). However, the
$C\wedge G\wedge G$ term in the supergravity action is not invariant under the
abelian gauge transformation, and a quantum gauge anomaly has to be used to
restore the gauge symmetry. This is the generalised Green-Schwarz
\cite{green84} mechanism described by Horava and Witten \cite{horava96-2}, and
it fixes the the expansion parameter,
\begin{equation}
\epsilon={1\over 4\pi}\left({\kappa\over4\pi}\right)^{2/3}.
\end{equation}
There are also gravitational anomalies which add $R\wedge R$ terms to the
boundary condition (\ref{cbc}). However, for the present we shall restrict
attention to the low curvature regime and ignore $R\wedge R$ terms.  

The remaining boundary conditions can be obtained by finding the extrema of the
action. Leaving the gauge multiplet fixed for the moment, the relevant boundary
terms in the variation of the action are
\begin{equation}
\delta S={2\over\kappa^2}\int_{\cal \partial M}d\mu
\left(\delta g_{AB}\,p^{AB}+\delta\bar\psi_i\,\theta^i
\right),\label{vars}
\end{equation}
with
\begin{eqnarray}
p^{AB}&=&-\frac12\left(K^{AB}-Kg^{AB}\right)+\frac14\kappa^2 T^{AB}
\label{pbc}\\
\theta^A&=&-\Gamma^{AB}P_+
\psi_B-\frac{\epsilon}{4}\Gamma^{BC}\Gamma^A\hat F^a{}_{BC}\chi^a
-\frac{\epsilon}{96}\Gamma^{AB}\Gamma^{CDE}\psi_B\bar\chi\Gamma_{CDE}\chi,
\end{eqnarray}
where $T^{AB}$ is the surface stress energy tensor.

The boundary condition $p^{AB}=0$ corresponds, in the covering space, to a
junction condition across the brane \cite{chamblin99}. The boundary condition
$\theta^A=0$ is a chirality condition on the gravitino
\begin{equation}
P_+\psi_A={\epsilon\over 12}\left(\Gamma_A{}^{BC}-10\delta_A{}^B\Gamma^C\right)
\hat F^a{}_{BC}\chi^a-\frac{\epsilon}{96}
\bar\chi\Gamma_{BCD}\chi \Gamma^{BCD}P_-\psi_A\label{fbc}
\end{equation}
This boundary condition represents a significant difference between the present
model and the construction of Horava and Witten \footnote{The
supersymmetry transformation of the chirality condition
$\Gamma_{11}\psi_A=\psi_A$ used by Horava and Witten \cite{horava96-2} in the
covering space approach is ambiguous because the supersymmetry transformation
of the gravitino is discontinuous across the brane.}. One significant
difference is that torsion can be generated by the matter fields. This affects
the connection $\hat\Omega$ which is governs the motion of gauginos.

The chirality condition on the gravitino and the boundary condition on the
three form play a special role in the supersymmetry of the theory.
The supersymmetry transformation rules are almost conventional,
\begin{eqnarray}
\delta g_{IJ}&=&\bar\eta\Gamma_{(I}\psi_{J)}\label{varg}\\
\delta\psi_I&=&D_I(\hat\Omega)\eta+
{\sqrt{2}\over 288}\left(\Gamma_I{}^{JKLM}-8\delta_I{}^J\Gamma^{KLM}\right)
\eta\hat G_{JKLM}\\
\delta C_{IJK}&=&-{\sqrt{2}\over 8}\bar\eta\Gamma_{[IJ}\psi_{K]}
+\frac{\sqrt{2}}4\epsilon\partial_{[I}f_{JK]}\\
\delta A^a{}_A&=&\frac12\bar\eta\Gamma_A\chi^a\\
P_-\delta \chi^a&=&-\frac14\Gamma^{AB}\hat F^a{}_{AB}\eta\label{vchi}
\end{eqnarray}
The two-form $f$ will be explained below. The restriction on the chirality of
the supersymmetry parameter is less conventional, but is determined by the
need to ensure supercovariance of the fermion boundary condition (\ref{fbc}),
\begin{equation}
P_+\eta=
-\frac{\epsilon}{96} \bar\chi\Gamma_{ABC}\chi\Gamma^{ABC}P_-\eta.
\end{equation}
This can be regarded as a modification of the projection operator $P_+$. The
chirality condition on the gaugino is simply $P_+\chi=0$ and is not affected by
any modification due to a convenient Fierz rearrangement.

We shall consider the supersymmetry of the boundary conditions first, and then
move on to the supersymmetry of the action. The variation of $C_{ABC}$
introduces a two-form $f$, which can be any two-form with the boundary
condition
\begin{equation}
f_{AB}=2A^a{}_{[A}\delta A^a{}_{B]}
\end{equation}
The two-form is needed to cancel a gauge non-invariant term from the variation
of the Chern Simons form in the boundary condition (\ref{cbc}). The remaining
terms in the variation of $C_{ABC}$ depend only on $P_+\psi_A$, allowing us to
use the gravitino boundary condition to show consistency with the variation of
the remaining terms. A calculation has been done to confirm that the boundary
condition (\ref{cbc}) is supersymmetric to {\em all} orders in $\epsilon$. A
more detailed discussion will be presented in a longer publication.

Taking the exterior derivative of $c_{ABC}=0$ and making use of (\ref{fbc})
puts the boundary condition in manifestly gauge invariant (and supercovariant)
form,
\begin{equation}
\hat G_{ABCD}=-3\sqrt{2}\epsilon\hat F^a{}_{[AB}\hat F^a{}_{AB]}
+\sqrt{2}\epsilon\bar\chi^a\Gamma_{[ABC}D_{D]}(\hat\Omega)\chi^a
+\frac{\sqrt{2}}{4}\epsilon\bar\chi^a\Gamma_{[ABC}\Gamma^{EF}\psi_{D]}\hat
F^a{}_{EF}.\label{gbc}
\end{equation}
This can be used to check the supersymmetry of the gravitino boundary condition
(\ref{fbc}). 

The variation of the action under the supersymmetry transformations can be
obtained by combining the boundary terms (\ref{vars}) with terms from the
variation of the matter multiplet and with terms which arise from the
interior. The invariance of 11-dimensional supergravity ensures that the
volume terms in the variation cancel. Boundary terms arise from the interior
when partial integration has to be used.

If we write the gravitino variation in the interior in the form
$\delta\psi_IL^I$, then integration by parts adds a term $\bar\eta L_N$ to the
other boundary terms in (\ref{vars}), where
\begin{equation}
 L_N=-\Gamma_N\Gamma^{AB}D_A(\hat\Omega)\psi_B
-\frac{\sqrt{2}}{96}\Gamma_N\Gamma^{ABCDE}\psi_A\hat G_{BCDE}
+\frac{\sqrt{2}}8\Gamma_N\Gamma^{AB}\psi^C\hat G_{ABCN}
\end{equation}
A slight rearrangement gives
\begin{eqnarray}
\bar\eta L_N&=&\bar\eta D_A(\hat\Omega)\theta^A+
\frac12K_{AC}\bar\eta\Gamma^{AB}\Gamma^C\psi_B
+\frac14\epsilon \bar\eta D_A(\hat\Omega)\left(\Gamma^{BC}\Gamma^A\hat
F_{BC}\chi\right)
\nonumber\\
&&-\frac{\sqrt{2}}{96}\bar\eta\Gamma^{ABCDE}\psi_A\hat G_{BCDE}
+\frac{\sqrt{2}}8\bar\eta\Gamma^{AB}\psi^C\hat G_{ABCN}.
\end{eqnarray}
We see clearly the term which cancels the extrinsic curvature in (\ref{pbc})
and a term involving the Yang-Mills supercurrent which partially cancels the
variation of the Yang-Mills part of the action. The penultimate term cancels
the variation of the supercurrent (using \ref{gbc}) and the final term
partially cancels with the variation of $C_{ABC}$. 

The three-form and gravitino boundary conditions
have to be imposed in order to obtain a supersymmetric action. After
some calculation, the action is found to be supersymmetric up to terms of
order $\epsilon^3$ (equivalent to $\kappa^2$).  At this order there are terms
arising from the variation of the $C\wedge G\wedge G$ term in
the action and no obvious source of terms for these to cancel. This is an
advance over previous attempts to construct the action which have encountered
infinite terms at order $\kappa^{4/3}$. 

The action and boundary conditions provide a supersymmetric theory which is a
natural candidate for the a low energy limit of $M$-theory. Further work is
needed to investigate fully what happens to the supersymmetry at order
$\kappa^2$ and to complete the supersymmetry variation of the gravitino
boundary condition. Going beyond the low energy limit, it would be desirable
to include the $R\wedge R$ curvature terms which are significant in
compactifications with Calabi-Yau manifolds
\cite{witten96,banks96,lukas98,lukas98-2,ellis99}.

\bibliography{super.bib}


\end{document}